\documentstyle[aps]{revtex}
\draft
\preprint{\begin{tabular}{l}SNUTP 98/036\\ HUTP-98/A024\end{tabular}}
\begin{document}
\title{\Large\bf Hidden Sector Gaugino Condensation and
the Model-independent Axion }
\author{Howard Georgi$^{(a)}$, Jihn E. Kim$^{(a,b,c)}$, and
Hans-Peter Nilles$^{(d,e)}$}
\address{$^{(a)}$Lyman Laboratory of Physics, Harvard University,
Cambridge, MA 02138\\
$^{(b)}$Department of Physics and Center for Theoretical Physics, 
Seoul National University, Seoul 151-742, Korea\\
$^(c)$ School of Physics, Korea Institute for Advanced Study,
207-43 Cheong-ryang-ri-dong, Seoul 130-012, Korea\\
$^{(d)}$Physikalisches Institut, Universit$\ddot a$t Bonn,
Nussallee 12, D-53115 Bonn, Germany\\
$^{(e)}$Max-Planck-Institut f\"ur Physik, D-80805 M\"unchen, Germany}
\maketitle

\begin{abstract}
In the effective field theory framework, we consider the effect of
supersymmetry breaking via gaugino condensation and supergravity in the
hidden sector gauge group on the hidden sector vacuum angle $\theta_h$.
The $\theta_h$ parameter dependence of the potential yields
phenomenologically acceptable invisible axion solutions if the $U(1)_R$
symmetry is broken down to a discrete subgroup $Z_N$ with $N\ge 3$ ($N=4$
is marginal).  Anomalous U(1) superstring models are good candidates for
this invisible axion resolution of the strong CP puzzle.
\end{abstract}
\newpage

The QCD vacuum angle $\bar\theta$ is bounded very strongly by the
upper bound on the neutron electric dipole moment to
$\bar\theta<10^{-9}$. The almost vanishing strong interaction
parameter has led to the so-called strong CP puzzle.
The axion \cite{reviews} may be the most attractive
solution of the strong CP puzzle, though other possibilities
such as the massless up quark or calculable $\bar\theta$ may
also be viable~\cite{choi,lewt,calc}.

However, the Peccei-Quinn (PQ) global symmetry \cite{pq}, the 
global symmetry needed for the axion solution, or the up quark
chiral symmetry is ad hoc in gauge theories,
since the gravitational interaction is expected to break all global
symmetries explicitly. Even if the axion solution is
attractive theoretically and cosmologically, one has to overcome
the puzzle of global symmetry in the
presence of the gravitational interaction.
Because superstring theory may provide a consistent approach to quantum
gravity, it seems reasonable to require that
any global symmetry that is introduced
be obtained in string theory.
In this regard, the discovery of the model-independent axion (MIa) in
superstring models \cite{witten} is of most fundamental importance.
This MIa suffers from the axion decay constant problem
\cite{ck}. On the other hand, this axion decay constant problem can be
circumvented with anomalous U(1) gauge groups \cite{kim}.

The argument of \cite{kim} is worth repeating.
Some 4D superstring models have an anomalous U(1)
\cite{dsw}, but the anomaly is not troublesome due to the Green-Schwarz
term \cite{green} which effectively introduces a coupling of the MIa with the
anomalous U(1) gauge boson
\begin{equation}
\epsilon_{MNOPQRSTUV}B^{MN}{\rm Tr}F^{OP}\langle F^{QR}\rangle
{\rm Tr}\langle F^{ST}\rangle\langle F^{UV}\rangle
\end{equation}
where the vacuum expectation value is of order the compactification
scale. In 4D, the coupling gives a quadratic term
$\sim M_c\partial^\mu a_{MI}A_\mu$ showing that the MIa ($a_{MI}$)
becomes the longitudinal component of the anomalous $U(1)$ gauge
boson $A_\mu$. This breaks the anomalous $U(1)$ gauge symmetry and the PQ
symmetry associated with translation of $a_{MI}$, leaving a global symmetry
below the compactification scale~\cite{kim}.

In more detail, the way this works is as follows. The original nonlinearly
realized tree level global symmetry is the symmetry,
\begin{equation}
\begin{array}{r@{\;}c@{\;}l}
a_{MI}&\rightarrow& a_{MI}+\alpha F_{MI}\\
\psi_i&\rightarrow& e^{iQ_ia_{MI}/F_{MI}}\,\psi_i
\end{array}
\label{nlgs}
\end{equation}
where $\alpha$ is a constant,
$F_{MI}$ is the MIa decay constant and
$\psi_i$ are the matter fields that transform nontrivially under the symmetry.
Because of (\ref{nlgs}), the MIa has only derivative interactions.

The anomalous $U(1)$ gauge symmetry takes the form
\begin{equation}
\begin{array}{r@{\;}c@{\;}l}
A_\mu&\rightarrow& A_\mu+(1/g_1)\partial_\mu\beta(x)\\
\psi_i&\rightarrow& e^{iq_i\beta(x)}
\psi_i\\
a_{MI}&\rightarrow&a_{MI}+M_c\,\beta(x)
\end{array}
\label{au1}
\end{equation}
where $M_c$ is the compactification scale, so that covariant derivatives have
the form
\begin{equation}
\begin{array}{r@{\;}c@{\;}l}
D_\mu\,\psi_i&=&\biggl(\partial_\mu
-ig_1q_iA_\mu\biggr)\,\psi_i\\
D_\mu\,a_{MI}&=&\partial_\mu\,a_{MI}-g_1M_c\,A_\mu
\end{array}
\end{equation}
Below the scale $M_c$, the MIa  and its interactions completely disappear,
because it is gauged away to become the longitudinal component of the
anomalous $U(1)$ gauge boson.
The heavy $A_\mu$ fields disappear from the
low energy theory, and
the original global PQ symmetry
is irrelevant at low energies. However,
there is an anomalous global symmetry in the low energy theory.
This is a linear combination of the PQ symmetry and a global anomalous gauge
symmetry, with $\beta(x)=-\alpha\,F_{MI}/M_c$.

Thus, below the compactification scale $M_c$, the MIa
does not exist, but there is a global $U(1)$ symmetry which we
will call $U(1)_X$.
The $U(1)_X$ is an anomalous global symmetry because
it involves the anomalous $U(1)$. This makes it an appropriate candidate to be
the PQ symmetry in the low energy theory, below $M_c$.

In spontaneously broken gauge models, it has long been known that
the symmetry breaking of a $U(1)$ gauge and a $U(1)$ global
symmetry by the VEV of one Higgs scalar can preserve a global
symmetry below the spontaneous symmetry breaking scale~\cite{hooft}.
The discussion above is in essence this 't~Hooft mechanism.

In models in which SUSY is broken by gaugino condensation in a hidden sector
and supergravity, we expect this condensation to spontaneously break the
$U(1)_X$ at a scale $F_X\sim 10^{12}$ GeV.
When the $U(1)_X$ breaks spontaneously, a new pseudoGoldstone boson is
produced
which we will call $a_X$. It is an axion, because the $U(1)_X$ symmetry is
anomalous. We emphasize that it is not the MIa --- it is ``made of'' the
hidden sector gaugino fields that condense to break $U(1)_X$ at the scale
$F_X$.\footnote{The translation of this effective field theory statement into
the language of the full field theory below the string scale is this.
Because there are contributions to the breaking of the two symmetries from
both of the two scales, $M_c$ and $F_X$, there are two Goldstone fields,
$a_{MI}$ and the ``invisible axion'', $a$, which is a linear combination of
phases of
$U(1)_X$ breaking singlet fields. One combination of $a_{MI}$ and
$a$, $a_A\simeq a_{MI}+\epsilon
a$ where $\epsilon\sim F_a/M_c$, is eaten to become the
longitudinal degree of the anomalous gauge boson.
The low energy invisible axion is the orthogonal linear combination,
$a_X\simeq a-\epsilon
a_{MI}$. It contains a small component of the model-independent
axion.}

If QCD were the only confining force, $a_X$ would be the perfect
choice for the invisible axion in superstring models. However,
we are assuming that the hidden sector that is required for
supersymmetry breaking involves
another confining interaction~\cite{gaugino}. Thus we have two
vacuum angles --- $\theta$ of QCD and $\theta_h$ of the hidden
sector confining gauge group. The $a_X$ contributes to both angles:
\begin{equation}
\theta={a_{X}\over F_{X}}+\theta^0,\ \ \theta_h={a_{X}\over F_{X}}
+\theta_h^0
\end{equation}
where $\theta^0$ and $\theta_h^0$ are parameters given by the
CP violations of weak interactions and hidden sector physics.
In general, $\theta_h^0\ne \theta^0$. Since the hidden sector
scale is much larger than the QCD scale, we might 
naively expect the hidden
sector interactions to give
a much larger contribution than QCD to the potential. These hidden sector
interactions are minimized for $\theta_h= 0$, and if they dominate, we will
find $a_{X}\simeq -F_{X}\theta_h^0$. But then the QCD angle $\theta$ is set to
$\theta^0-\theta_h^0$, not to zero, and the strong CP puzzle is not resolved.
To set both $\theta$ and $\theta_h$ to zero, we would need two independent
axions. But it seems that except for the global symmetry
related to the MIa $U(1)_{MI}$, there is no bosonic global
symmetry in superstring models~\cite{banks}.

Thus to implement the invisible axion
with a confining hidden sector, we must find theories in which
the $\theta_h$ dependence of the vacuum energy is suppressed. 
In fact, some suppression is built into supersymmetric theories of this kind 
because the bare gaugino mass is much smaller than the scale $F_X$ of gaugino
condensation, because it arises only due to the supersymmetry breaking
produced by the supergravitational interactions. However, this is not enough
to supress the $\theta_h$ dependence to acceptable levels.

   If the hidden sector gluino carried an unbroken anomalous $U(1)$ symmetry
like the PQ symmetry in a model with a massless $u$ quark, that would
eliminate the $\theta_h$ dependence of the vacuum energy entirely and
allow the QCD interactions to determine the vacuum value of $a_{X}$ and set
$\theta=0$.\footnote{And, of course, it would ensure that the mass term for
the hidden sector gaugino remained zero even after SUSY breaking.} A $U(1)_R$
symmetry rotates the hidden sector gaugino fields and is therefore a
candidate for such a symmetry. However, we do not expect any such global
continuous symmetry to be exact in the presence of gravitational interactions.
Thus the best we can do is to find models in which the contribution of the
hidden sector to the $a_{X}$ potential is sufficiently suppressed, because of
the precise form of the $U(1)_R$ breaking interactions. For this purpose, we
look for models with the following properties:

\indent (i) A $Z_N$ subgroup of the continuous $U(1)_R$ symmetry survives even
in the presence of the gravitational effects that explicitly break the
continuous $U(1)_R$,\footnote{This may be possible in string theory even if
there are no continuous global symmetries.} and\\
\indent (ii) The spontaneous $U(1)_R$ violation at the scale $F_X$ occurs only
through hidden sector gaugino condensation and supergravity.

Let us now examine the consequence of these assumptions. Viable superstring
models would lead to effective low energy supersymmetry with
nonrenormalizable terms suppressed by string or Planck scale.
The dominant low energy physics from a D=4 supergravity theory is
characterized by three functions, the
superpotential, the K\"ahler potential, and
the gauge kinetic function,
\begin{equation}
W(\phi_i),\ \ K(\phi_i,\phi^*_i),\ \ f_1^a(\phi_i)
\end{equation}
where $\{\phi_i\}$ is a set of chiral superfields, and the
subscript $a$ of the gauge kinetic function runs over different
gauge groups, $G=\prod_a G_a$.
Supersymmetry is broken when
$\langle G^i\rangle\ne 0$ where $G=K+\log|W|^2$ and
$G^i=\partial G/\partial\phi_i$ (from here on, we use gravitational units in
which the Planck mass is one). The gravitino mass
becomes $m_{3/2}=e^{G/2}=e^{K/2}|W|$.

In fact, both the superpotential and
\begin{equation}
f_1^a(\phi_i)\,W^{a\alpha} W^a_\alpha
\end{equation}
(where $W^a_\alpha$ is the gauge field strength superfield for the group $a$)
depend only on left-handed superfields, and their $F$ terms appear in the
Lagrangian.\footnote{Because are primarily interested in gaugino
condensation, we consider only $f$'s that are gauge singlets --- otherwise
there would be more general terms here, as below in (\ref{nterms}).} 
But if gaugino condensation occurs at a very large scale, we must
also consider similar terms in the Lagrangian with higher powers of
$W^{a\alpha} W^a_\alpha$,
where $a$ refers to the hidden sector gauge group, because below the
condensation scale, the powers of $W^{a\alpha} W^a_\alpha$ are replaced by 
powers of $\Lambda_h^3$, where $\Lambda_h$ is the large hidden sector scale.
These terms can be parameterized as follows:
\begin{equation}
\int d^2\theta \sum_{a,n}f^a_n(\phi_i)(W^{a\alpha} W^a_\alpha)^n
\label{nterms} 
\end{equation}
where the $f^a_n$ functions are holomorphic in $\phi$.
Because of our assumption (ii), these are the terms that give the dominant
contribution to $R$-symmetry breaking in the low energy theory.

Furthermore, because the spontaneous $R$ symmetry breaking in the hidden
sector is assumed to arise from
gaugino condensation only, the $R$ breaking terms from (\ref{nterms}) in the
low energy theory are
evaluated at zero VEV of chiral fields (assumption (ii) again).
We can take 
\begin{equation}
f^a_n=\epsilon^a_n S
\label{nepsilon}
\end{equation}
where
$S$ is a chiral superfield. Terms with more powers of chiral
fields, such as $S^l (l>1), SA^nB^m\cdots (n,m\ge 1)$, etc.
do not contribute since we assumed VEV's of chiral fields
are vanishing. By power counting, the $\epsilon$ parameters in
(\ref{nepsilon}) are suppressed by
appropriate powers of the Planck mass, $M_P$. In general, there could be
a whole set of $S$ fields, but the effect of one will be enough to
illustrate our mechanism.

From (\ref{nterms}) and (\ref{nepsilon}), we get contributions to the
potential energy of the form
\begin{equation}
V=\left|\left[(\lambda\cdot\lambda)+\sum_{n\ge 2}
\epsilon^a_n(\lambda\cdot\lambda)^n\right] \right|^2
\label{vr}
\end{equation}
where the $\lambda$s are the hidden sector
gaugino fields and where we have normalized the coefficient of the first term
\begin{equation}
\epsilon^a_1=1.
\end{equation}
If only the first term exists, i.e. $f^a_n=0$ for $n\ge 2$,
then (\ref{vr}) is not $R$ violating.\footnote{Another way to see this is that
we could choose the $R$ quantum number of $S$ to preserve $R$ invariance.}

Let us now consider the effect of our assumption that a $Z_N$ subgroup of
$U(1)_R$ remains unbroken. Under such a $Z_N$ subgroup,
\begin{equation}
\lambda\rightarrow e^{2\pi i/N}\lambda
\label{zn}
\end{equation}
Thus if $N$ is even, $(\lambda\lambda)^{jN/2}$ is invariant, for any integer
$j$, and if $N$ is odd $(\lambda\lambda)^{jN}$ is invariant. In either case,
the sum over $n$ in (\ref{vr}) will be restricted.
The most important $R$-violating
term in the potential will result from the first allowed term in the sum over
$n$, and will have the form
\begin{equation}
\epsilon_m(\bar\lambda\cdot\bar\lambda)(\lambda\cdot\lambda)^m
+{\rm h.c.}
\end{equation}
where $m$ is the lowest integer greater than 1 for which $\epsilon_m\neq0$,
which is $N+1$ for $N$ odd, and ${N\over2}+1$ for $N$ even.
The interesting values of $m$ are $m\geq3$. The $m=2$ ($N=2$) case is not
interesting because in that case, the $Z_2$ symmetry does not forbid the
gaugino mass term.

Notice that the $Z_N$ symmetry must be a subgroup of
the continuous $R$ symmetry because the
gauge kinetic term, $\int
d^2\theta W^\alpha W_\alpha$ must be invariant --- 
the gauginos and the gauge bosons in the multiplet must transform differently.

Our question is what is the
effect of the above form of $R$ symmetry breaking on
the dependence on the hidden sector parameter $\theta_h$.
To consider this problem, we take an effective field theory
viewpoint below the hidden sector scale $\Lambda_h$. The light
degree of freedom arising from the gaugino condensation is the
pseudoscalar meson $\eta^\prime$ whose mass is about
$\Lambda_h$. The parameters describing the effective theory are
$\epsilon_m$, $\theta_h$, and the scale for the gaugino
condensation. Since this phenomenon is very similar to
the chiral symmetry breaking of QCD, we first recall some basic
facts about the $\theta$ dependence in QCD.

As the simplest example, let us consider up quark condensation
in one-flavor QCD. The mass term in this theory is given by
\begin{eqnarray}
{\cal L}_{mass} = - m_u \bar{u} u
\end{eqnarray}
Formally, we can assign the following $U(1)$ chiral transformation,
\begin{eqnarray}
u &\longrightarrow& e^{i\alpha} u\nonumber\\
\bar{u} &\longrightarrow& e^{i\alpha} \bar{u}\nonumber\\
m &\longrightarrow& e^{-2i\alpha} m\\
\theta &\longrightarrow& \theta + 2\alpha \nonumber
\end{eqnarray}
From the above chiral symmetry, we expect the following effective potential
below the chiral symmetry breaking scale,
\begin{eqnarray}
&V = \frac{1}{2} m_u \Lambda^3 e^{i\theta} - \frac{1}{2}\lambda_1
  \Lambda v^3 e^{i \frac{\eta}{v} - i\theta}
    -\frac{1}{2}\lambda_2 m_u v^3 e^{i \frac{\eta}{v}}
  +\lambda_3 m_u^2\Lambda^2e^{2i\theta}\nonumber\\
&+\lambda_4{v^6\over \Lambda^2}e^{2i{\eta\over v}-2i\theta}
  +\cdots + \mbox{h.c.}
\end{eqnarray}
where $\cdots$ denotes higher order terms,
$\lambda$'s are couplings of  order 1, $\langle\bar{u} u\rangle =
v^3 e^{i \eta/v}$, and the QCD scale $\Lambda$ is inserted
to make up the correct dimension.\footnote{Note that we are not trying to keep
careful track of the powers of $2\pi$ here, so the distinction between the
scale $\Lambda$ and the amplitude for $\eta$ production $v$ has not meen made
very precisely.} In addition, $e^{\pm i\theta},
e^{\pm 2i\theta}$, etc is multiplied to respect the $U(1)$ symmetry.
Note that if $m_u \neq 0$ and $\theta$
is not a dynamical variable, then the strong CP puzzle is not solved.
However, if $m_u = 0$ then only the $m_u$-independent terms survive,
leading to
\begin{eqnarray}
  V = -{1\over 2}\lambda_1 \Lambda v^3 e^{i \frac{\eta}{v} - i\theta}
  +\lambda_4 {v^6\over\Lambda^2}e^{2i{\eta\over v}-2i\theta}
  +\cdots+\mbox{h.c.}
\end{eqnarray}
Thus, redefining the $\eta$ field as $\eta^{\prime}$
\begin{eqnarray}
 \eta^{\prime} = \eta - v \theta,
\end{eqnarray}
the $\theta$ dependence is completely removed from $V$. The $\theta$
parameter is unphysical if a quark is massless. As is well
known the massless up quark scenario solves the
strong CP puzzle even though it obtains
a constituent quark mass when the chiral symmetry is
broken. Of course, as we discussed in the introduction, the relevance of this
solution to the strong CP puzzle hinges on
its viability in hadron physics phenomenology \cite{lewt}.

To compare with the gaugino condensation case, we comment
on the possible violation of the chiral symmetry by gravitational
interactions for the case of the massless up quark. Even if the
mass term is not present, chiral symmetry breaking
nonrenormalizable interactions may be generated,
\begin{equation}
{1\over M_P^2}\bar uu\bar uu e^{-2i\theta},
{1\over M_P}\bar uu \sigma\sigma e^{-i\theta},\ \
{\rm etc.}
\end{equation}
where $\sigma$ is the singlet field carrying no chiral charge.
The first term will shift $\theta$ by a tiny amount $|\langle \bar uu
\rangle|^{2/3}/M_P^2\sim 10^{-38}$ which is ridiculously small. But if
some scalar singlets develop a large expectation value, one
expects a sizable $\theta\sim |\langle\sigma\rangle|^2/M_P
|\langle\bar uu\rangle|^{1/3}$.
The upper bound on $\theta$ implies $|\langle\sigma\rangle|<10^{4}$ GeV.

In the same way, the spontaneous $R$ symmetry breaking by the
hidden sector gaugino condensation dictates the $\theta_h$ parameter
dependence of the potential through the explicit $R$ breaking
$\epsilon_m$ dependence. As in QCD, we can study the form
of the potential by treating $\epsilon_m$, as a spurion field, and
imposing the symmetry
\begin{eqnarray}
&\lambda\rightarrow e^{i\alpha}\lambda\nonumber\\
&\epsilon_m\rightarrow e^{-2mi\alpha}\epsilon_m\\
&\theta_h\rightarrow\theta_h+2\ell (G)\alpha\nonumber
\end{eqnarray}
where $\ell (G)$ is the index of the adjoint representation of the
gauge group $G$. We assume the hidden sector group as $SU(K)$ for
which $\ell =K$. The effective interaction is
\begin{equation}
V_{\rm eff}={1\over 2}\bar\lambda\bar\lambda\lambda\lambda
+\epsilon_m\bar\lambda\bar\lambda(\lambda\lambda)^m
+{\Lambda_h^4\over\Lambda_h^{3K}}
\,(\lambda\lambda)^Ke^{-i\theta_h}
+\cdots+{\rm h.c.}
\end{equation}
which includes the terms
\begin{equation}
\left(v\over M_P\right)^{3m-1}v^4 e^{i(m-1)\eta/v}
+\left(v\over \Lambda_h\right)^{3K}\Lambda_h^4
e^{i({K\eta\over v}-\theta_h)}+\cdots+{\rm h.c.}
\label{someterms}
\end{equation}
where we have replaced $\epsilon_m$ by the appropriate number of powers of
$M_P$.
Redefining $\eta^\prime=\eta-v\theta_h/K$ (because the second term in
(\ref{someterms}) is much larger than the first and enforces this
identification), we obtain the $\theta_h$
dependence
\begin{equation}
{v^{3m+3}\over M_P^{3m-1}}e^{-i(m-1)(\theta_h/K+\eta^\prime/v)}+\cdots+{\rm
h.c.}
\label{dthetah}
\end{equation}
This is the dominant term depending on $\theta_h$.

Let us now discuss the $\theta_h$ dependence in (\ref{dthetah}), remembering
that $m$ is the lowest integer greater than 1 for which $\epsilon_m\neq0$.
Clearly, the larger $m$, the smaller the $\theta_h$ dependence. Even if the
hidden sector scale is
large, $\Lambda_h\sim 10^{12}$ GeV, the height of the $\theta_h$
potential is very low. 
The situation is summarized in table 1. 

For $N=3$ the supression is adequate over almost all of the range.
$N=4$ is marginal. But all higher $N$ provide adequate supression.

\begin{center}
\begin{minipage}[t]{25em}{\begin{center}
\begin{tabular}{|ccccc|}
\hline
$Z_N$ &\ & $\epsilon_m$ &\ & $V\; {\rm(GeV)^4}$\\
\hline
$N$=3 &\ &\ \ \ $m$=4 &\ & $\ \ \sim 10^{-29}-10^{-8}$\\
$N$=4 &\ &\ \ \ $m$=3 &\ & $\ \ \sim 10^{-8}-10^{10}$\\
$N$=5 &\ &\ \ \ $m$=6 &\ & $\ \ \sim 10^{-50}-10^{-26}$\\
\hline
\end{tabular}\\ \end{center}
\vspace{1.5em}
\noindent
Table 1: The $\theta_h$ dependence of potential in
GeV$^4$ units for $\Lambda_h=10^{12}-10^{13}$ GeV
and Planck mass suppression factor of $10^{19}-
2.4\times 10^{18}$ GeV.
}
\end{minipage}\end{center}
\vspace{3em}

In conclusion, we have studied the implication of the hidden sector
gaugino condensation on breaking of $Z_N$ subgroups of $U(1)_R$.
Under the assumption that $U(1)_R$ breaking VEV's of chiral
fields do not exist, we showed that anomalous $U(1)$
models with $N\ge 3$ ($N=4$ is marginal) are viable candidates
containing phenomenologically acceptable invisible axion models.

In string models, discrete
symmetries like the $Z_N$ we posit arise naturally. 
In fact, the discrete symmetries
were used before~\cite{lps} to obtain an accidental PQ symmetry.
Our idea differs from Ref.~\cite{lps} in that in our case
the PQ symmetry is present in anomalous $U(1)$ gauge models
and we employ a specific discrete symmetry, i.e.
the discrete subgroup of $U(1)_R$, to suppress the
hidden sector contributions to the axion mass. In Ref.~\cite{lps},
the PQ symmetry arises accidentally and the extra confining group is
not considered.

\acknowledgments
The work of HG and JEK is supported in part by the Distinguished
Scholar Exchange Program of Korea Research Foundation and NSF-PHY 92-18167.
JEK is also supported in part by the Hoam Foundation, and the work
of HPN is partially supported by grants from TMR programs
ERBFMRX--CT96--0045 and CT96--0090.

\end{document}